\documentclass[12pt]{article}
\usepackage{graphicx}
\usepackage{lmodern}
\usepackage{caption}
\usepackage{amsmath}
\usepackage{authblk}
\usepackage{booktabs}

\usepackage{amsfonts}
\usepackage{authblk}
\usepackage[T1]{fontenc}

\usepackage{hyperref}

\usepackage{colortbl}
\usepackage[dvipsnames]{xcolor}

\begin{document}
\renewcommand\Affilfont{\fontsize{10}{10.8}\itshape}
\newcommand{\lang}{(}
\newcommand{\rang}{)}

\newcommand{\fig}{.}

\newcommand{\cc}{\cellcolor{Cerulean!10}}

\title{{\em Multi-SpaM}: a Maximum-Likelihood approach to Phylogeny Reconstruction based on {\em Multi}ple {\em Spa}ced-Word {\em M}atches}
\author[1]{Thomas Dencker}
\author[1]{Chris-Andre Leimeister}
\author[2]{Michael Gerth}
\author[3,4]{Christoph Bleidorn}
\author[5]{Sagi Snir}
\author[1,6]{Burkhard Morgenstern}

\affil[1]{University of G\"ottingen, 
 Department of Bioinformatics,
Goldschmidtstr. 1, 37077 G\"ottingen, Germany}
\affil[2]{Institute for Integrative Biology, University of Liverpool, Biosciences Building, Crown Street, L69 7ZB Liverpool, UK}
\affil[3]{University of G\"ottingen, 
 Department of Animal Evolution and Biodiversity,
Untere Karsp{\"u}le 2, 37073 G\"ottingen, Germany}
\affil[4]{Museo Nacional de Ciencias Naturales,
Spanish National Research Council (CSIC), 28006 Madrid, Spain}
\affil[5]{Institute of Evolution,
Department of Evolutionary and  Environmental Biology, University of Haifa,
199 Aba Khoushy Ave.  Mount Carmel, Haifa, Israel}
\affil[6]{G{\"o}ttingen Center of Molecular Biosciences (GZMB),
Justus-von-Liebig-Weg 11, 37077 G{\"o}ttingen, Germany}

\maketitle

\begin{abstract}
\textbf{Motivation:} 
Word-based or `alignment-free' methods for phylogeny
reconstruction are much faster
than traditional approaches, but they are generally less
accurate. Most of these methods calculate pairwise distances for a set
of input sequences, for example from {\em word frequencies}, from so-called {\em spaced-word matches} or from the average {\em length of common substrings}.
\\ 
\textbf{Results:}
In this paper, we propose the first word-based approach to tree reconstruction that is based on multiple sequence comparison and {\em Maximum Likelihood}.
Our algorithm first samples small, gap-free
alignments involving four taxa each. For each of these alignments,
it then calculates a quartet tree and, finally, the program {\em Quartet MaxCut} is used to infer a super tree topology for the full set of input
taxa from the calculated quartet trees. Experimental results show that trees
calculated with our approach are of high quality.\\
\textbf{Availability:} 
The source code of the program is available at\\ 
\href{\tt https://github.com/tdencker/multi-SpaM}{\tt https://github.com/tdencker/multi-SpaM} \\  
\textbf{Contact:} \href{thomas.dencker@stud.uni-goettingen.de}{thomas.dencker@stud.uni-goettingen.de}\\
\end{abstract}

\section{Introduction}

To gain a better understanding of the evolution of genes or species, reconstructing accurate phylogenetic trees is essential. This can be done using standard methods which rely on  \textit{sequence alignments}, either of entire genomes or of sets of orthologous genes or proteins.    
\textit{Character-based} methods such as {\em Maximum Parsimony} \cite{far:70,fit:71} or {\em Maximum Likelihood} \cite{fel:81} infer trees based on evolutionary substitution events that 
may have happened since the species evolved from a common ancestor. 
These methods are generally considered to be  accurate, 
as long as the underlying alignments are of high quality, {and as long
as suitable substitution models are used}. However,
for the task of multiple alignment no exact polynomial-time algorithm 
exists, and even heuristic approaches can be time consuming~\cite{sie:wil:din:etal:11}. 
Moreover, 
the most popular heuristic for multiple alignment, the {\em progressive
alignment} \cite{fen:doo:87}, has been shown to be relatively unstable: 
multiple alignments calculated with  
progressive approaches and trees inferred from these alignments  
depend on the underlying {\em guide trees} and even on the
order of the input sequences \cite{cha:flo:tom:etal:18}.
In addition to these difficulties, 
exact algorithms for {character-based} phylogeny approaches  
are themselves 
{\em NP hard}~\cite{cho:tul:05,AAM:FouldsG1982}.

{\em Distance} methods, by contrast, {infer phylogenies by estimating}  evolutionary distances for all pairs of input taxa \cite{fel:84}. Here,  
 pairwise alignments are sufficient which can be faster calculated than multiple alignments, but still require runtime proportional to the product of the lengths of the aligned sequences. There is a loss in accuracy, however, compared to character-based approaches, as all of the information about evolutionary events is reduced to a single number for each pair of taxa, 
and not more than two sequences are considered simultaneously, as opposed to character-based approaches, where all sequences are examined simultaneously. 
%
%
The final trees 
 are obtained by clustering based on distance matrices, most commonly with \textit{Neighbor Joining}~\cite{sai:nei:87}.
Since both pairwise and multiple sequence alignments are computationally expensive, they are ill-suited for the increasingly large datasets that are available today due to the next generation sequencing techniques.

In recent years, a number of \textit{alignment-free} approaches to genome-based phylogeny reconstruction have been published which are very fast in comparison to alignment-based methods~\cite{son:ren:rei:etal:14,zie:vin:alm:kar:17,ber:cha:rag:16,bro:gri:otw:16,piz:16,ren:bai:you:etal:18}. 
Another advantage of these new methods is that they circumvent some  well-known problems in genome alignment such as genome rearrangements and duplications. Moreover, alignment-free methods can be applied 
to incomplete sequence sets and even to collections of unassembled reads \cite{roy:vis:bha:13,son:ren:zha:etal:13,yi:jin:13,com:sch:14}.
 A disadvantage of these methods is that they are, in general, considerably less accurate than slower, alignment-based methods. 


Some `alignment-free' approaches compare fixed-length {\em words} of the input sequences to each other, so   
-- despite being called `alignment-free' -- they are using local pairwise `mini-alignments'.  
Recently, methods have been proposed that estimate phylogenetic distances based on the relative frequency of mismatches in such local alignments.  
An example is \textit{co-phylog}~\cite{yi:jin:13} which finds short gap-free alignments of a fixed length, consisting of matching nucleotide pairs only, except for the middle position where a mismatch is allowed.  
Phylogenetic distances are estimated from the fraction of such alignments for which the middle position is a mismatch. As a generalization of this approach, \textit{andi}~\cite{hau:klo:pfa:14} uses pairs of maximal exact word matches that have the same distance to each other in both sequences; the frequency  of mismatches in the segments between those matches is then used to  estimate the number of substitutions per position between two input sequences. 

{\em co-phylog} and {\em andi} require a minimum length of the flanking word matches in order to reduce the number of random background matches. Threfore, they tend not to perform well on distantly related sequences where long exact matches are less frequent. 
Moreover, the number of random segment matches grows quadratically with the length of the input sequences while the expected number of homologous matches grows only linearly. Thus, longer exact matches are necessary in these approaches to limit the number of background matches if longer sequences are compared. This, in turn, reduces the number of {homologies} that are found, and therefore the amount of information that can be used to calculate accurate distances.
Other alignment-free approaches are based on the length of maximal common substrings between sequences. These approaches are also very efficient, since  common substrings can be rapidly found using suffix trees or related data structures \cite{uli:bur:tul:cho:06,hau:pfa:dom:wie:09}.  As a generalization of this approach, some methods use longest common substrings with a certain number of mismatches \cite{lei:mor:14,tha:apo:alu:16,tha:cho:liu:etal:17,mor:sch:lei:17,aya:cha:ili:pis:18}.

Recently, we proposed to use words with {\em wildcard characters} -- so-called \textit{spaced words} -- for alignment-free sequence comparison \cite{lei:bod:hor:lin:mor:14,hor:lin:bod:etal:14}. 
Here, a  binary pattern of {\em match} and {\em don't-care} positions specifies the positions of the {\em wildcard} characters \cite{oun:lon:15,noe:17,gir:com:piz:18}. In \textit{Filtered Spaced-Word Matches (FSWM)}~\cite{lei:soh:mor:17} and {\em Proteome-based Spaced-word Matches (Prot-SpaM)} \cite{lei:sch:sch:etal:18}, alignments of such spaced words are used, where sequence positions must match at the {\em match} positions while mismatches are allowed at the \textit{don't care positions}. 
A score is calculated for every such spaced-word match in order to remove -- or {\em filter out} -- {\em background} spaced-word matches; the mismatch frequency of the remaining \textit{homologous} spaced-word matches is then used to estimate the number of substitutions per position that happened since two sequences evolved from their last common ancestor. The filtering step allows us to use patterns with fewer match positions in comparison to above mentioned methods {\em co-phylog} and {\em andi}, since the vast majority of the background noise can be eliminated reliably by looking at the {\em don't-care} positions of the initially found spaced-word matches. As a result, the phylogenetic distances calculated by {\em FSMW} and {\em Prot-SpaM} are generally rather accurate, even for large and distantly related sequences.

In this paper, we introduce a novel approach to phylogeny reconstruction called {\em \underline{\smash{Multi}}ple \underline{\smash{Spa}}ced-Word \underline{\smash{M}}atches (Multi-SpaM)} that combines the {\em speed} of the so-called `alignment-free' methods with the 
{\em accuracy} of the  {\em Maximum-Likelihood} approach. While other alignment-free methods are limited to {\em pairwise} sequence comparison, we generalize the above outlined {{\em spaced-word} approach to {\em multiple} sequence comparison. For a binary pattern of {\em match} and {\em don't care} positions, {\em Multi-SpaM} identifies {\em quartet blocks} of four matching spaced words each, {\em i.e.} gap-free four-way alignments with matching nucleotides at the {\em match} positions of the underlying  binary pattern and possible mismatches at the {\em don't care} positions, see Figure~\ref{fig:block} for an example. 
For each such quartet block, an optimal {\em Maximum-Likelihood} tree topology is calculated with the software {\em RAxML}~\cite{sta:14}.
The \textit{Quartet MaxCut} algorithm~\cite{sni:rao:12} is then used to combine the calculated quartet tree topologies into a super tree. 
We show that on both simulated and real data, {\em Multi-SpaM} produces phylogenetic trees of high quality and often outperforms other alignment-free methods.  

\section{Method}
\subsection{Spaced words and $P$-blocks} 
To describe our method, we first need some formal definitions. A \textit{spaced word} of length~$\ell$ exists in the context of a binary pattern $P \in \{0,1\}^\ell$ of the same length. This pattern marks every position as either a \textit{match position} in case of a 1 or as a \textit{don't care position} in case of a 0. The number of match positions is called the \textit{weight} of the pattern.
Given such a pattern~$P$, a \textit{spaced word}~$w$ is a word of length~$\ell$ over the alphabet \{A,C,G,T,*\} such that $w(i) = *$ if and only if $P(i) = 0$, {\em i.e.} if and only if $i$ is a {\em don't care} position. The symbol `$*$' is interpreted as a  `wildcard' character. 
For a DNA Sequence $S$ of length $n$ and a position $0 \leq i \leq n-l+1$, we say that a \textit{spaced word} $w$ occurs in $S$ at position~$i$ -- or that  $[S,i]$ is an {\em occurrence} of $w$ -- if $S(i + j) = w(j)$ for all match positions~$j$.
This follows the definition that we previously used \cite{lei:bod:hor:lin:mor:14,mor:zhu:hor:lei:15}.

A pair $\left([S,i],[S',i']\right)$ of occurrences of the same spaced word~$w$ is called a {\em spaced-word match}. For a substitution matrix assigning a {\em score} $s\lang X,Y\rang$ to every pair $(X,Y)$ of nucleotides, we define the {\em score} of a spaced word match $([S,i],[S',i'])$ as  
$$\sum_{P(k) = 0}  s\lang S(i+k),S'(i'+k)\rang$$
That is, if we align the two occurrences of $w$ to each other, the score of the spaced-word match is the sum of the scores of the nucleotides aligned to each other at the {\em don't-care} positions of~$P$. In {\em Multi-SpaM}, we are using the  nucleotide substitution matrix below that has been proposed by
Chiaromonte {\em et al.}~\cite{chi:yap:mil:02}:

\[
\begin{array}{crrrr}
  & A  &   C  &  G  &   T  \\
A & 91 & -114 & -31 & -123 \\
C &    & 100  &-125 &  -31 \\
G &    &      & 100 & -114 \\
T &    &      &     &   91 \\
\end{array}
\]

{\em Multi-SpaM} starts with generating a binary pattern~$P$ with  user-defined length~$\ell$ and weight~$w$. By default, we use values $\ell=110$ and $w=10$, {\em i.e.} by default the pattern has 10 {\em match positions} and 100 {\em don't-care} positions, but other values for $\ell$ and $w$ can be chosen by the user. Given these parameters, $P$ is calculated by running our previously developed software tool {\em rasbhari}~\cite{hah:lei:oun:etal:16}.  

As a basis for phylogeny reconstruction, we are using four-way alignments consisting of occurrences of the same spaced word with respect to $P$ in four different sequences. We call such an alignment a {\em quartet $P$-block} or a $P$-{\em block}, for short. A $P$-block is thus a gap-free alignment of length $\ell$ where in the $k$-th column identical nucleotides are aligned if $k$ is a {\em  match} position in $P$, while mismatches are possible if $k$ is a {\em don't-care} position, see Figure~\ref{fig:block} for an example.   
\begin{figure}
\begin{center}
$\begin{array}{rlllllllllllll}
S_0: & T & A &  \cc \bf{C} & \cc\bf{T} & \cc A & \cc \bf{G} & \cc C & \cc G & \cc \bf{T} & C & G &  & \\
S_1: & A & C & T & C & \cc \bf{C} & \cc \bf{T} & \cc A & \cc \bf{G} & \cc T & \cc G & \cc \bf{T} & T & G  \\
&&&&&&&&&&&&&\\
\end{array}$
\end{center}
\caption{
\label{fig:sw-match}
Spaced-word match with respect to a pattern $P=1101001$. The score 
of this spaced-word match would be $s\lang A,A\rang + s\lang C,T\rang 
+ s\lang G,G\rang = 91 - 31 + 100 = 160.$  
}
\end{figure}
Note that the number of such $P$-blocks can be very large: if there are $n$ occurrences of a spaced-word~$w$ in $n$ different sequences, then this gives rise to $n \choose 4$ different $P$-blocks. 
Thus, instead of using all possible $P$-blocks, {\em Multi-SpaM} randomly samples a limited number of $P$-blocks  to keep the program runtime under control.  

Moreover, for phylogeny reconstruction, we want to use $P$-blocks that are likely to represent true homologies. Therefore, we introduce the following definition: a $P$-block -- {\em i.e.} a set of four occurrences of the same spaced word~$w$ --  is called a {\em homologous $P$-block} if it contains at least {\em one} occurrence $[S_i,p]$ of $w$  such that all remaining three occurrences of $w$  have positive scores when compared to $[S_i,p]$.  
To sample a list of homologous $P$-blocks, we randomly select spaced-word occurrences with respect to~$P$ from the input sequences and their reverse complements. For each selected $[S_i,p]$, we then randomly select occurrences of the same spaced word from sequences $S_j \not= S_i$, until we have found three occurrences of $w$ from three different sequences that all have positive scores with $[S_i,p]$.
\begin{figure}
\begin{center}
$\begin{array}{rlllllllllllll}
S_{0}: & C & \cc \bf{C} & \cc \bf{C} & \cc A & \cc A \cc & \cc \bf{G} & G & A & C & &&& \\
S_{1}: & A & A & C & T & A & C & G & T & A & C & C & T &  \\
S_{2}: & A & A & C & T & A & C & G & T & A & C & C & &  \\
S_{3}: & \cc \bf{C} & \cc\bf{C} & \cc A & \cc C & \cc \bf{G} & T & C & C & G & C & G &&\\
S_{4}: & A & G & A & C & T & C & \cc \bf{C} & \cc \bf{C} & \cc A & \cc A & \cc \bf{G} & G & A  \\
S_{5}: & T & C & \cc \bf{C} & \cc \bf{C} & \cc A & \cc T & \cc \bf{G} & G & A & C & C && \\
S_{6}: & A & A & C & T & A & C & G & T & A & C & C & A &  \\
&&&&&&&&&&&&&\\
       & 0 & 1 & 2 & 3 & 4 & 5 & 6 & 7 & 8 & 9 & 10 & 11 & 12 \\
\end{array}$
\end{center}
\caption{
\label{fig:block}
$P$-block for a pattern $P=11001$: the spaced word $w= {CC}**G$ occurs
at $[S_0,1], [S_3,0], [S_4,6]$ and $[S_5,2]$.}  
\end{figure}

To find spaced-word matches efficiently  we
first sort the list of all occurrences of spaced words with respect to $P$ in lexicographic order. This way, we obtain a list of spaced-word occurrences where all occurrences of the same spaced word $w$ are appearing as a contiguous block. 
Once we have sampled a homologous $P$-block as described, we remove the four occurrences of $w$ from our list of spaced-word occurrences, so no two of the sampled $P$-blocks can contain the same occurrence of a spaced word. 
The algorithm continues to sample $P$-blocks until no further $P$-blocks can be found, or until a given number 
of  $P$-blocks is reached. By default, {\em Multi-SpaM} uses a maximal number of $M=1,000,000$ $P$-blocks, but this parameter can be adjusted by the user.  

\subsection{Quartet trees}

{For each of the sampled quartet $P$-blocks, we infer an unrooted tree topology. This most basic {\em unrooted} phylogenetic unit is called a \textit{quartet} topology; there are three possible different quartet
topologies for a set of four taxa.  
To identify the best of these three topologies, we use the {\em Maximum Likelihood} program {\em RAxML}~\cite{sta:14}.  We note that 
{\em RAxML} is a general {\em Maximum-Likelihood} software, its use in our context is fairly degenerated, as we only use it to 
infer optimal quartet topologies.}\\
\begin{figure}[h]
\centering
\includegraphics[width=0.5\linewidth]{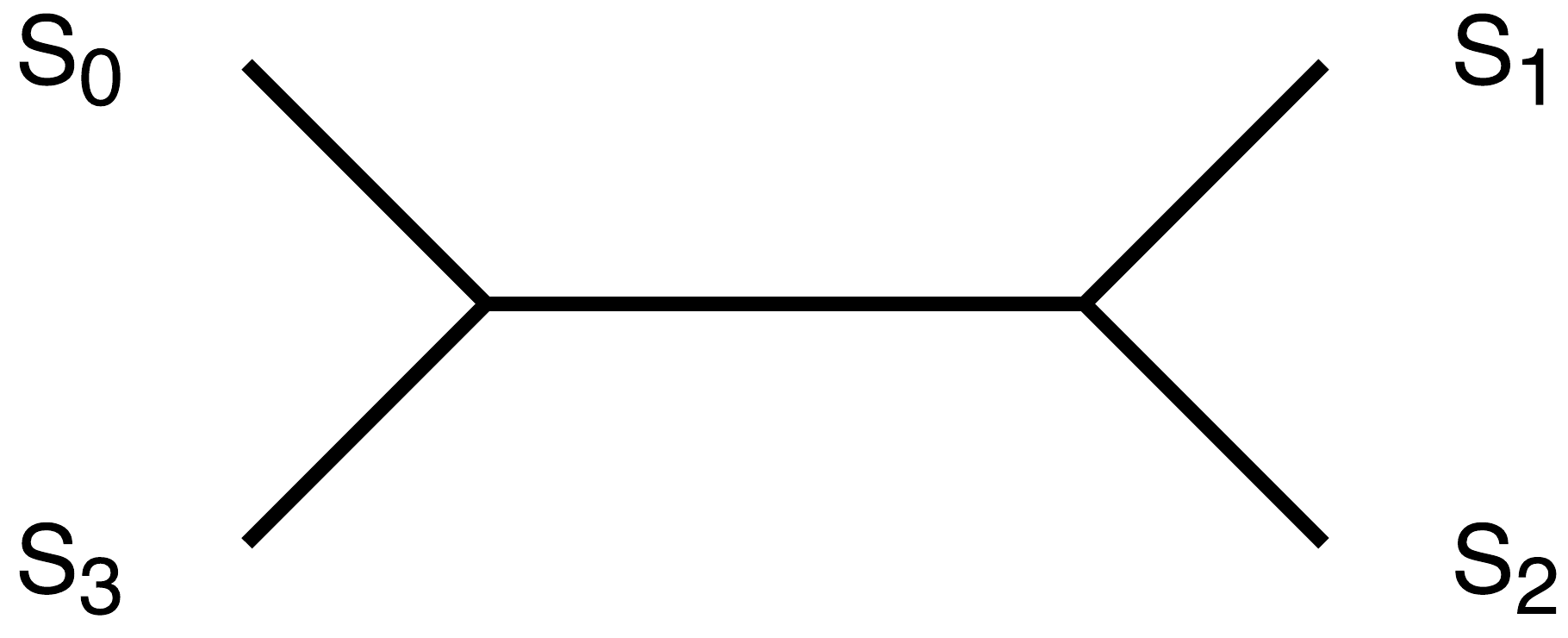}
\label{fig:qrtt}
\caption{Example of a quartet tree topology.}
\end{figure}

After having the optimal tree topology for each of the sampled quartet P-blocks, we need to amalgamate  them into a single tree spanning the entire taxa set. This task is denoted the {\em Supertree Task}~\cite{bininda-book-2004}
and is known to be {\em NP hard}, even for the special case where the input 
is limited to 
quartets topologies, as in our case~\cite{NPsteel}. Nevertheless there are several heuristics
for this task, with {\em MRP}~\cite{Baum1992,Ragan1992} the most popular. Here we chose to use {\em Quartet MaxCut}~\cite{sni:rao:10,sni:rao:12}
that proved to be faster and more accurate for this kind of input~\cite{Avni-JME-2018}. In brief, {\em Quartet MaxCut} partitions recursively the taxa set where
each such partition corresponds to a split in the final tree. In each
such recursive step, a graph over the taxa set is built where the set
of quartets induces the edge set in that graph. The idea is to
partition the vertex set (the taxa) such that the minimum quartets are
violated. This is achieved by a {\em semidefinite-programming}-like
algorithm that embeds the graph on the unit sphere and applies a
random hyperplane through the sphere.

\subsection{Implementation}
To keep the runtime of our software manageable, we integrated the {\em RAxML} code directly into our program code.  
We parallelized  our program with {\em openmp}~\cite{openmp:02}. 

\section{Test results}
To evaluate {\em Multi-SpaM}, we applied it to both simulated and real sequence data and compared the resulting trees to reference trees. In phylogeny reconstruction, artificial benchmark data are often  used since here, `correct' reference trees are known. For the real-world sequence data that we used in our study, we had to rely on reference trees that are believed to {reflect the true evolutionary history}, or on trees calculated using traditional, alignment-based methods that can be considered to be reasonably accurate. In our test runs, we used standard parameters for all methods, if such parameters were suggested by the respective program authors. The program {\em kmacs} that was one of the programs that we evaluated,  has no default value for its only parameter, the number $k$ of allowed mismatches in common substrings. Here, we chose a value of $k = 4$. 
While {\em Multi-SpaM} produces tree topologies without branch lengths, 
all other methods that we compared, produce distance matrices. To generate trees with these methods, we applied {\em Neighbor-joining}~\cite{sai:nei:87} to the distances produced by these methods.  

\begin{figure}[h]
\includegraphics[width=\linewidth]{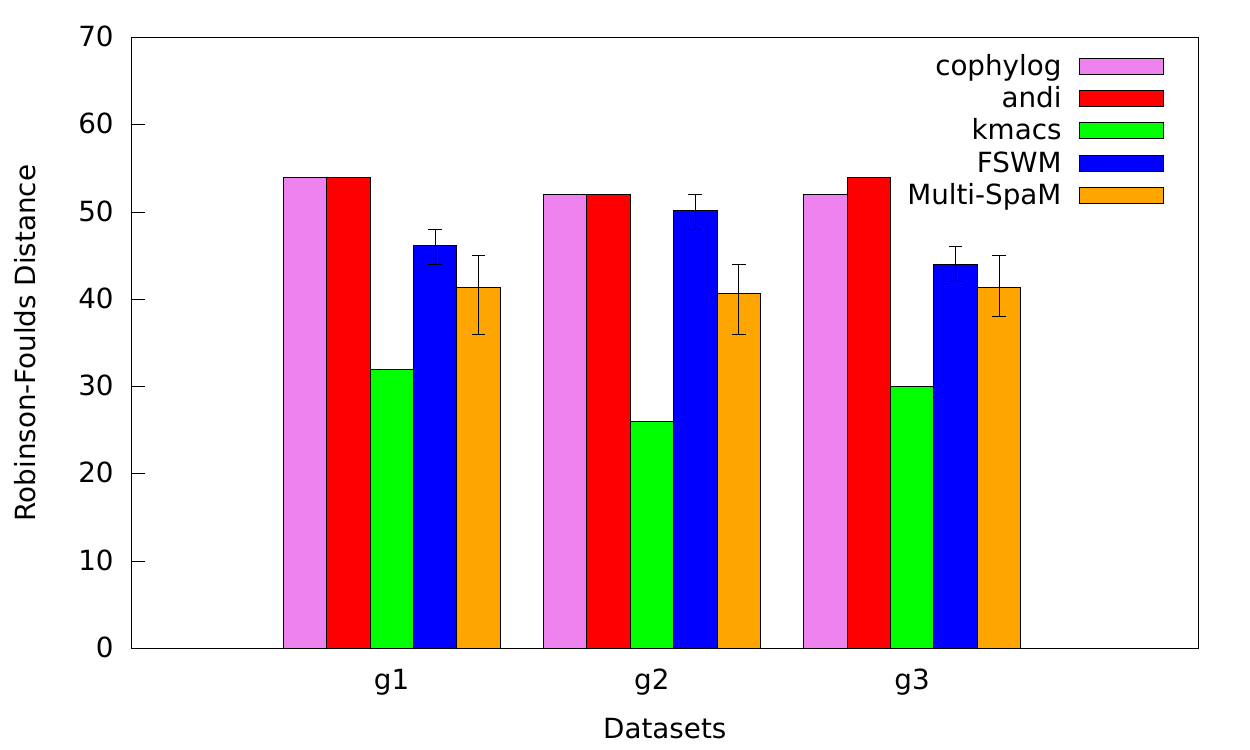}
\caption{Average {\em Robinson-Foulds (RF)} distances between trees calculated
with alignment-free methods 
and reference trees for three datasets of simulated bacterial genomes. {\em FSWM} and {\em Multi-SpaM} were run 10 times, with different patterns~$P$ generated (see main text). Error bars indicate the lowest and highest {\em RF} distances, respectively.}
\label{fig:bac}
\end{figure}

\begin{figure}[h]
\includegraphics[width=\linewidth]{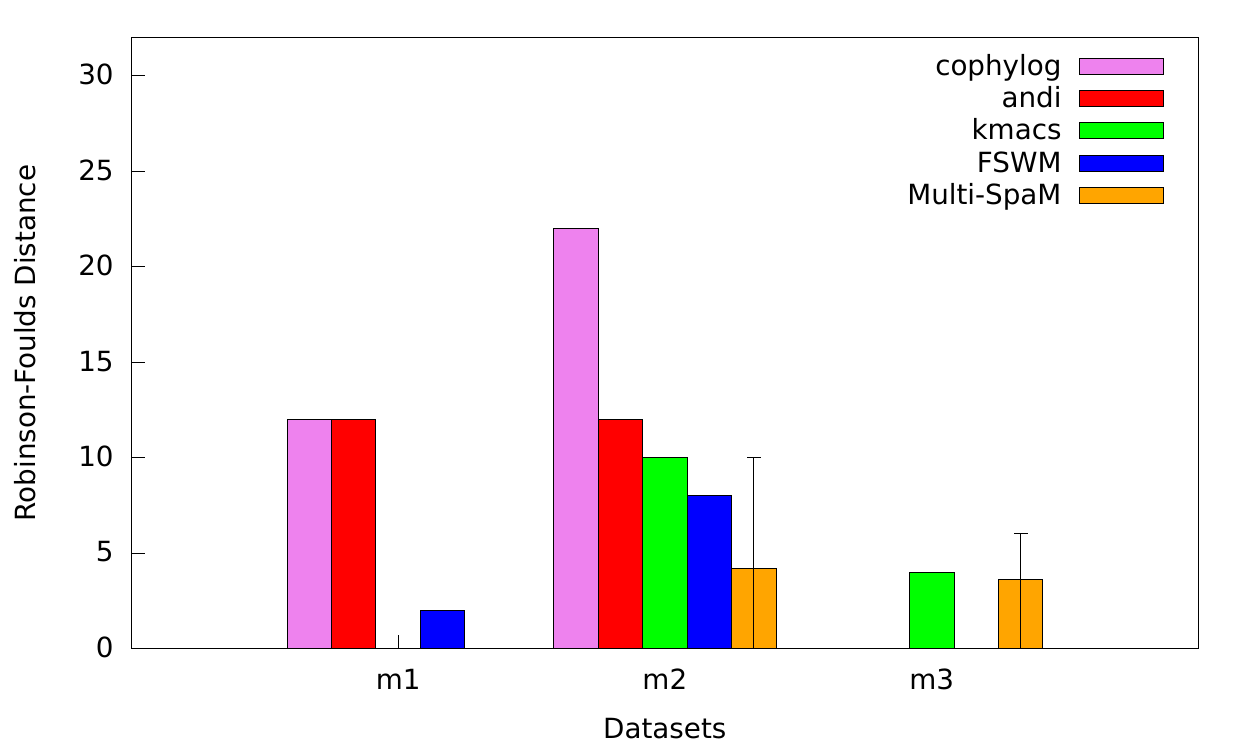}
\caption{{\em RF} distances for three sets of simulated mammalian genomes. If no bar is shown, the {\em RF} distance is zero for the respective method and data set. {\em E.g.} the {\em RF} distance between the tree generated by {
\em kmacs} for data set {\em m1} and the reference tree is zero, {\em i.e.} here the reference tree topology was precisely reconstructed. Error bars for {\em FSWM} and {\em Multi-SpaM} are as in Figure~\ref{fig:bac}.}
\end{figure}

To compare the trees produced by the different alignment-free methods to the respective benchmark trees, we used the  {\em Robinson-Foulds (RF)} metric~\cite{rob:fou:81}, a standard measure to compare how different two tree topologies are. Thus, the smaller the {\em RF} distances between reconstructed trees and the corresponding reference trees are, the better a method is. 
For the {\em RF} metric as well as for {\em Neighbor-joining}, we used the {\em PHYLIP} package~\cite{fel:89}.
As explained above, both {\em FSWM} and {\em Multi-SpaM} rely on  binary patterns of {\em match} and {\em don't care} positions. The results of these programs therefore depend on the underlying patterns. Both programs use the software {\em rasbhari}~\cite{hah:lei:oun:etal:16} to calculate patterns. {\em rasbhari} uses a probabilistic algorithm, so different program runs usually return different patterns and, as a result, different program runs with {\em FSWM} and {\em Multi-SpaM} may produce slightly different distance estimates, even if the same parameter setting is used. To see how {\em FSWM} and {\em Multi-SpaM} depend on the underlying binary patterns, we ran both program ten times on each data set. The figures in the {\em Results} section report the {\em average}  
{\em RF}-distance for each data set over the ten program runs. Error bars indicate the highest and lowest {\em RF}-distances, respectively.

\subsection{Simulated Sequences}

At first, we evaluated {\em Multi-SpaM} on datasets generated with the {\em Artificial Life Framework (ALF)}~\cite{dal:ani:gon:des:12}. {\em ALF} starts by  simulating an ancestral  genome that includes a number of genes. According to a guide tree which is either provided by the user or randomly generated, {\em ALF} simulates speciation events and other evolutionary events such as substitutions, insertions and deletions for nucleotides as well as duplications and deletions of entire genes. A large number of parameters can be specified by the user for these events.
We used parameter files that were used in a study by the authors of {\em ALF}~\cite{dal:ani:gon:des:12}. This way, we generated six datasets, three with simulated  bacterial genomes, and three with simulated mammalian genomes. 
We used the base parameter sets for each dataset and only slightly modified them to generate {\em DNA} sequences for roughly 1,000 genes per taxon which we then concatenated to full genomes. The size of the data set  is around 10~{\em mb} each.
As shown in Figure~\ref{fig:bac}, none of the tools that we evaluated were able to exactly reconstruct the reference tree topologies for the simulated bacterial genomes. 
In contrast, reference topologies for the simulated mammalian genomes could be reconstructed by some tools, although no method could reconstruct all three reference topologies exactly.

\subsection{Real genomes}

We also applied the programs that we evaluated to real genomes to see if the results are similar to our results on simulated genomes. Here, our first dataset were 29 {\em E.~coli/Shigella} genomes which are commonly used as a benchmark dataset to evaluate alignment-free methods~\cite{hau:klo:pfa:14}. As a reference, we used a tree calculated with {\em Maximum Likelihood}, based on a \textit{mugsy} alignment~\cite{ang:sal:11}. The dataset is 144~{\em mb} large and the average distance between two sequences in this set is about 0.0166 substitutions per sequence position.
Next, we used a set of 32 {\em Roseobacter} genomes of 132 {\em mb} with a 
reference tree published by Newton {\em et al.}~\cite{Newton2010}; here the 
distance between sequence pairs was 0.233 substitutions per position on 
average.    
 As a third benchmark set, we used 
 19 {\em Wolbachia} genomes that have been analyzed by Gerth and Bleidorn~\cite{ger:ble:16}; we used the phylogeny published in their paper as a reference. 
The total size of this sequence set is 25 {\em mb}, 
the average pairwise distance is 0.06 substitutions per position. 
The results of these three series of test runs are summarized in Figure~\ref{fig:real}. 
On the {\em E. coli / Shigella} data set, our new approach  was outperformed
by three out of four competing methods that we evaluated. 
On the other two data sets, 
{\em Roseobacter} and {\em Wolbachia}, however, {\em Multi-SpaM}  turned
out to be the best performing method.

\begin{figure}[h]
\includegraphics[width=\linewidth]{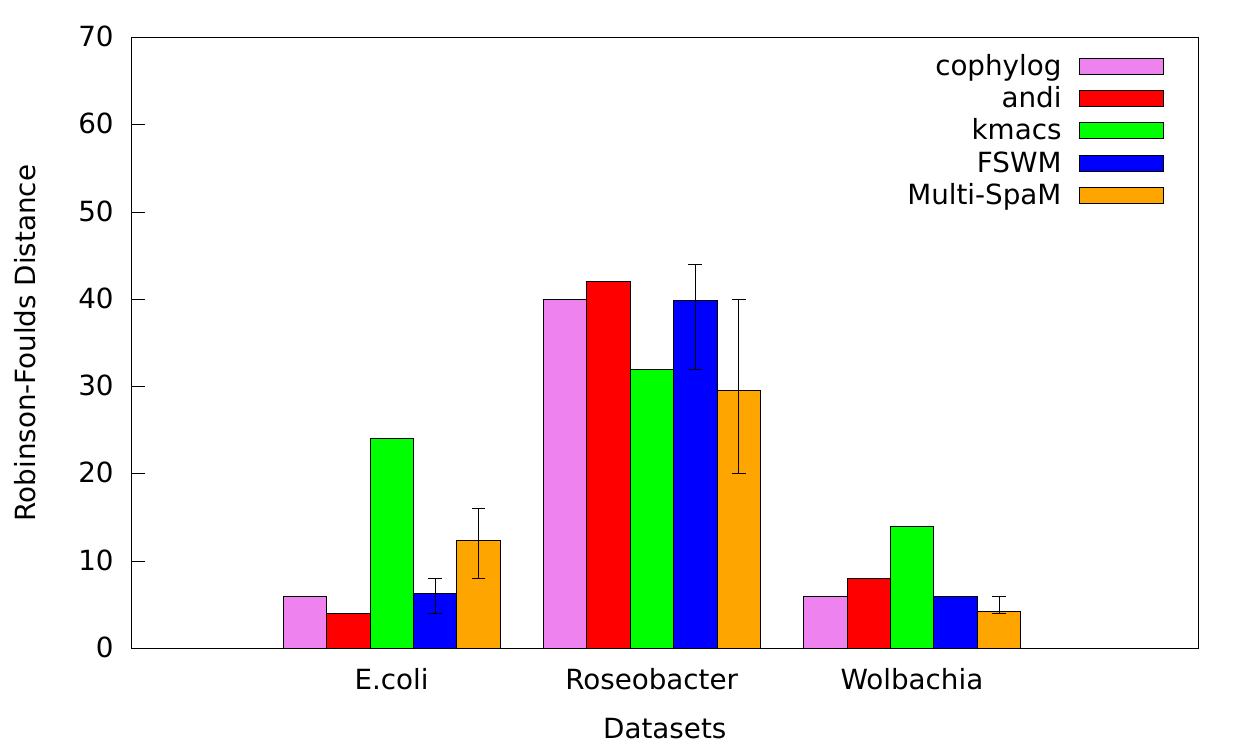}
\caption{{\em RF} distances for three sets of benchmark genomes: 
29 {\em E.~coli/Shigella} genomes, 32 {\em Roseobacter} genomes and 
 19 {\em Wolbachia} genomes.
 Error bars for {\em FSWM} and {\em Multi-SpaM} as in Figure~\ref{fig:bac}.}
\label{fig:real}
\end{figure}

Finally, we applied our dataset to a much larger dataset of eukaryotic genomes. It consists of 14 plant genomes totalling 4.8 {\em GB}. Figure~\ref{fig:plants} shows the resulting trees.  
For this data set, we used a pattern with a weight of $w=12$ instead of the default value $w=10$, to keep the number of background spaced-word matches manageable. For a dataset of this size, the number of additional score calculations would increase the runtime unnecessarily if one would use the default  weight of $w=10$.  
As can be seen in Figure~\ref{fig:plants}, {\em Multi-SpaM} and {\em FSWM} produced fairly accurate trees for this data set, with only minor differences to the reference tree:
{\em Multi-SpaM} misclassified {\em Carica papaya}, whereas {\em FSWM} failed to classify {\em Brassica rapa} correctly. None of the other alignment-free tools that we evaluated  could produce a reasonable tree for this data set: {\em andi} returned a tree that is rather different to the reference tree, while {\em kmacs} and {\em co-phylog} could not finish the program runs in a reasonable timeframe.

As explained in the {\em Method} section, {\em Multi-SpaM} calculates an optimal tree topology for each of the sampled {\em quartet $P$-blocks}. Here, it  can happen that no single best topology is found. In particular for closely related sequences, this happens for a large fraction of the sampled quartet $P$-blocks. For the {\em E.~coli/Shigella} data set, for example,  
around 50\% of the quartet blocks were inconclusive, {\em i.e.} {\em RAxML} could find no single best tree topology. We observed a similar result for a dataset of 13 {\em Brucella} genomes where the pairwise phylogenetic distances are even smaller than for the {\em E.~coli/Shigella} data set, namely 0.0019  substitutions per site, on average. Here, roughly 80\% of the blocks were inconclusive. 
For all other datasets, the fraction of inconclusive quartet $P$-blocks was negligible.
For example, for the set of 14 plant genomes, only $\sim 250$ out of the 1,000,000 sampled $P$-blocks were inconclusive.  


\begin{figure*}
\includegraphics[width=\textwidth]{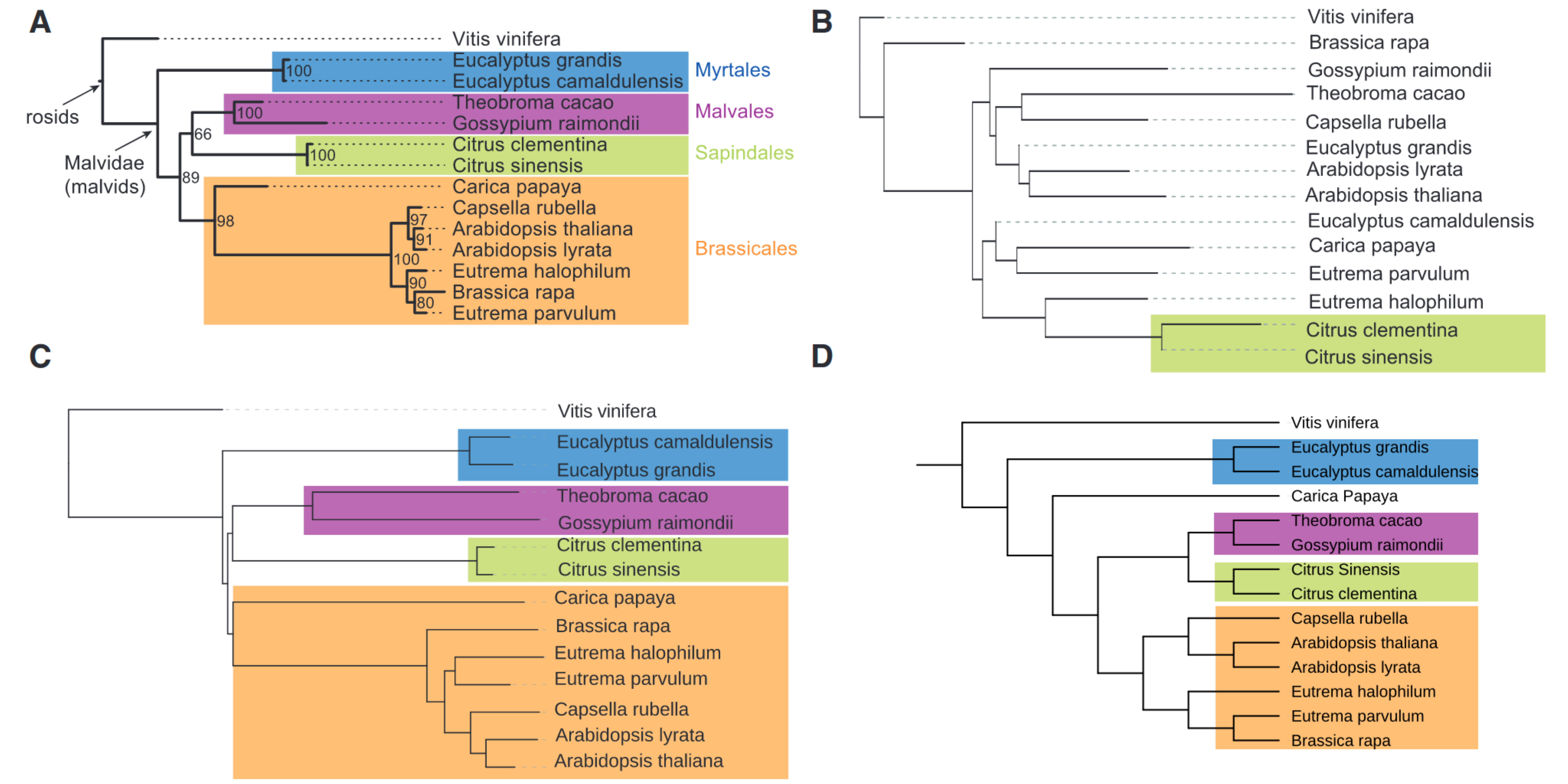}
\caption{Reference tree ({\bf A}) from Hatje and Kollmar \cite{hat:kol:12} and trees reconstructed by {\em andi} (\textbf{B}), {\em FSWM} (\textbf{C}) and  {\em Multi-SpaM} (\textbf{D}) for a set of 14 plant genomes.
\label{fig:plants}}
\end{figure*}

\subsection{Program runtime and memory usage}

Table \ref{time:tab} shows the program runtime for {\em Multi-SpaM, FSWM, kmacs, andi} and {\em co-phylog} on the three real-world data sets in our 
program comparison.  
The test runs were done on a 5 x Intel(R) Xeon(R) CPU E7-4850 with 2.00 GHz, a total of 40 threads (20 cores). 
For the largest data set in our study, the set of 14 plant genomes, the peak {\em RAM} usage was 76 {\em GB} for {\em FSWM}, 110 {\em GB} for {\em andi} 
and 142 {\em GB} for {\em Multi-SpaM}. In memory saving mode, the peak {\em RAM}
usage of {\em Multi-SpaM} could be reduced to 10.5 {\em GB}, but this 
roughly doubles the program runtime.

\begin{table}[h]
\centering
\begin{tabular}{lrrrrr}
\toprule
                & \em M.-SpaM & \em FSWM  & \em kmacs & \em andi & \em co-phylog \\
\midrule
\em E.coli/Shig.
                &    683         &      906  & 41,336    &   11     &  443          \\
\em Roseobac.   &  7,991         &      746  & 13,163    &   17     &  615          \\
\em Wolbachia   &    382         &       70  & 17,581    &    2.6   &   58          \\  
Plants          & 27,072         & 1,107,720 &  - \ \ \  & 1,808    & - \ \         \\
\bottomrule
\end{tabular}
\vspace*{0.5 cm} 
\caption{Runtime in seconds for different alignment-free approaches  on 
the four sets of real-world genomes that we used in this study. 
On the largest data set, the 14 plant genomes, {\em kmacs} and {\em co-phylog} did not terminate the program run. 
On this data set, we increased the pattern weight for {\em Multi-SpaM} 
from the default value of $w=10$ to $w=12$, in order to
reduce the runtime. Note that {\em Multi-SpaM}, {\em FSWM} and {\em andi}  are parallelized, so we could run them on
multiple processors, while {\em kmacs} and {\em co-phylog} had to be run on single processors. The reported
runtimes are {\em wall-clock} times. \label{time:tab}}
\end{table}

\section{Discussion}

Standard methods for phylogeny reconstruction are computationally expensive, 
because they rely on multiple sequence alignments and on time-con\-sum\-ing 
probabilistic calculations. Moreover, the number of possible trees to be compared under optimality criteria such as {\em likelihood} and {\em parsimony} grows exponentially with the number of taxa/sequences \cite{fel:04}.  
In contrast, existing 
alignment-free phylogeny methods are {\em distance based} approaches. 
They are generally regarded to be less accurate than {\em character-based} approaches, but are orders of magnitudes faster. 

In this paper, we introduced a novel approach to phylogeny reconstruction called {\em Multi-SpaM} to combine the speed of alignment-free methods with the accuracy of {\em Maximum Likelihood}.  
{To our knowledge, this is the first alignment-free approach that uses multiple sequence comparison and likelihood.}  
Our test runs show that {\em Multi-SpaM} produces phylogenetic trees of high quality. It outperforms other alignment-free methods, in particular on sequence sets with large evolutionary distances. For closely related input sequences, such as  different strains of the same bacterial species, however, our approach was sometimes outperformed by other alignment-free methods. As shown in Figure~\ref{fig:real}, the programs {\em andi, co-phylog} and {\em FSWM} produce better results on a set of {\em E.~coli/Shigella} genomes than {\em Multi-SpaM}. This may be due to our above mentioned observation that for many {\em quartet $P$-blocks} no single best tree topology can be found if the compared sequences are very similar to each other.

Calculating optimal tree topologies for the sampled {\em quartet $P$-blocks} is a relatively time-consuming step in {\em Multi-SpaM}. In fact, we observed that the program runtime is roughly proportional to the number of {\em quartet blocks}  for which  topologies are calculated.  However, the maximal number of {\em quartet blocks} that are sampled is a user-defined parameter. By default we sample up to $M=1,000,000$ quartet blocks; in our test runs the quality of the resulting trees could not significantly improved by increasing $M$. Consequently, our method is relatively fast on large data sets, where only a small fraction of the possible quartet-blocks is sampled. By contrast, on small data sets, {\em Multi-SpaM} is slower than other alignment-free methods.    
To improve the runtime of our approach, we have parallelized our software to run on multiple cores; the runtimes
in Table~\ref{time:tab} are wall-clock runtimes. 
It should be straight-forward to adapt our software to be run on distributed 
systems, as has been done for other alignment-free approaches
 \cite{cat:pet:gia:ros:17,pet:gue:piz:17}.


Apart from the maximum number of sampled quartet blocks, the only important parameters of our approach are the {\em length} and the {\em weight} --  {\em i.e.} number of {\em match positions} -- of the underlying binary pattern. For {\em Multi-SpaM}, we used similar default parameter values as in {\em Filtered Spaced Word Matches (FSWM)}, namely a  weight of $w=10$ and a pattern length of $\ell=110$, {\em i.e.} our default patterns have 10 {\em match} positions and 100 {\em don't-care} positions. A large number of {\em don't care} positions is important in {\em FSWM} 
and {\em Multi-SpaM} as this makes it easier to distinguish homologous from random background spaced-word matches. Also, with a large number of  {\em don't-care}  positions, the number of {\em inconclusive} quartet $P$-blocks is reduced for sequences with a high degree of similarity. On the other hand, we found that the number of {\em match} positions has less impact on the performance of {\em Multi-SpaM}.    
On large data sets it is advisable to increase the {\em weight} of~$P$ since this reduces the fraction of background spaced-word matches, and therefore the number of spaced words for which their scores need to be calculated by the program. A higher weight, thus, reduces the program runtime.  
For the largest data set our study, the set of plant genomes, we increased the pattern weight from the default value of 10 to  a value of  12 to keep the runtime of  {\em Multi-SpaM} low.

Our current implementation uses the previously developed software {\em Quartet MaxCut} by Snir and Rao \cite{sni:rao:10,sni:rao:12} to calculate supertrees from quartet tree topologies. We are using this program since it is faster than other supertree approaches.  As a result, the current 
version of {\em Multi-SpaM} generates tree {\em topologies} only, {\em i.e.} trees without branch lengths. We will investigate in the future, if our approach can be extended to calculate full phylogenetic trees with branch lengths, based on the same {\em quartet} $M$-blocks that we have used in the present study. 


\section*{Funding}
The project was funded by the {\em VW Foundation}, project
{\em VWZN3157}.

%
\bibliographystyle{plain}
%

\bibliography{multi-spam.bbl} 

\end{document}